\documentclass[aps,prb,citeautoscript,twocolumn,floatfix,showpacs]{revtex4}
\usepackage{amsmath,amssymb,graphics}
\usepackage{array,multirow}
\begin{document}
\title{Structural (B1 $\to$ B8) Phase Transition in MnO under
  Pressure: Comparison of All-electron and Pseudopotential Approaches}

\author{Jind\v rich Koloren\v c}
\altaffiliation[On leave from ]{Institute of Physics, Academy of
  Sciences of the Czech Republic, Na Slovance~2, CZ-18221 Praha~8,
  Czech Republic}
\author{Lubos Mitas}
\affiliation{Department of Physics, North Carolina State University,
  Raleigh, North Carolina 27695, USA}

\date{\today}

\begin{abstract}
We employ the density functional theory to study a structural
transition of MnO from B1 (rocksalt) to B8 (NiAs) structures that was
observed experimentally at pressures around 100~GPa. We utilize
all-electron description as well as norm-conserving pseudopotentials
and demonstrate that these two approaches can significantly differ in
quantitative predictions. We explicitly show 
that even small-core pseudopotentials exhibit transferability
inaccuracies for quantities sensitive to the energy differences
between high- and low-spin polarizations of valence electrons.
\end{abstract}


% 62.50.+p High-pressure and shock wave effects in solids and liquids
% 64.30.+t Equations of state of specific substances
% 64.70.Kb Solid-solid transitions
% 71.10.-w Theories and models of many-electron systems
% 71.15.-m Methods of electronic structure calculations
% 71.15.Nc Total energy and cohesive energy calculations
% 71.20.-b Electron density of states and band structure of crystalline solids
% 71.27.+a Strongly correlated electron systems; heavy fermions
% 71.30.+h Metal-insulator transitions and other electronic transitions
% 72.80.Ga Transition-metal compounds
% 75.10.Lp Band and itinerant models

\maketitle

\section{Introduction}

Transition metal oxides have received a considerable attention of
condensed matter physicists in the past several decades. One of the
reasons for such an elevated interest is that the standard band theory
methods, such as the density functional theory (DFT) represented by
the local density approximation (LDA) or by its semi-local
generalization (GGA), do not satisfactorily describe these
materials. For example, the LDA/GGA predicts incorrect equilibrium
crystal structure for FeO~\cite{mazin1998,fang1998}.  In the case of
equilibrium properties of MnO, these approximations perform reasonably
well as was shown by several groups \cite{pask2001,terakura1984}.  The
source of deficiencies of simple band theories is agreed to be
underestimation of correlations among $3d$ electrons. Improved account
for these correlations is needed to adequately capture Mott insulating
mechanism and the competing charge transfer effects.

Several methods have been devised to resolve the issue of strongly
correlated electrons, such as GW, %\cite{faleev2004}
LDA+U, %\cite{anisimov1991}
self-interaction correction (SIC) %\cite{svane1990}
or hybrid exchange-correlation functionals that combine
local exchange from LDA or GGA with non-local Fock
exchange. %\cite{feng2004,franchini2005}
Performance of several of these approaches has recently been
thoroughly compared in Ref.~~\onlinecite{kasinathan2006}, using the
pressure-induced high-spin to low-spin transition \cite{cohen1997} in
MnO as a benchmark test. It turned out that although the current
computational techniques work well for some aspect of the electronic
structure (such as the equilibrium volume or the character of the
ground state at ambient pressure), they are not quite dependable if
one considers high-pressure properties of this material.

One of the goals of this paper is to provide additional benchmarks for
the electronic structure of manganese oxide at high pressure,
expanding thus the test set of Ref.~~\onlinecite{kasinathan2006}.  Our
main message does not directly relate to the $3d$ electrons but to a
rather surprising finding that the deep core states and the type of
method used for their description can significantly influence certain
properties generally considered to be related to the valence electrons
only. An example of such a quantity is the critical pressure $P_c$ for
the pressure-induced transition from B1 (rocksalt) to B8 (NiAs)
structure that has been reported recently
\cite{kondo2000,yoo2005}. This transition is accompanied by a collapse
of the local magnetic moments of Mn atoms \cite{yoo2005} as well as
with electronic transition to a metallic state \cite{patterson2004}.

Another key motivation for this study of core-valence interactions in
crystalline MnO is our interest in applications of quantum Monte Carlo
(QMC) techniques to transition metal oxides and other materials
containing strongly correlated electrons.  The accurate representation
of physical ions by norm-conserving pseudopotentials is essential for
obtaining meaningful and predictive quantum Monte Carlo results and it
is therefore crucial to assess the accuracy and limits of such
pseudopotential Hamiltonians. Ideally, one would compare results of
pseudopotential QMC with all-electron QMC. Unfortunately, such a route
is prohibitively computationally expensive at present. We have
therefore decided to carry out a careful study within the DFT,
Hartree-Fock (HF) approximation and combination of these methods,
since the accuracy of one-particle orbitals is important in QMC as
mentioned below and explained elsewhere
\cite{wagner2003,wagner2006}.  Similar problems were investigated in
several density-functional studies (e.g.,
Refs.~~\onlinecite{louie1982,cho1996,kresse1999,kiejna2006}) and
provided important signals that the choice of the methods and
treatment of pseudopotentials can affect the final results.

Our study provides new insights in several important and complementary
directions. First, we employ Dirac-Fock pseudopotentials, which
nominally should not require any additional corrections (such as the
DFT nonlinear core correction as commented later on).  The reason for
this choice is that these pseudopotentials were designed to be
accurate for correlated wave function calculations and, in fact,
several tests have been carried out within both post-Hartree-Fock
(configuration interaction) and quantum Monte Carlo methods with quite
satisfactory results \cite{wagner2003,wagner2006}. Second, we wanted
to quantify the pseudopotential vs.~all-electron differences within
the DFT, HF and hybrid functionals since our previous
\cite{wagner2003} (and current preliminary) calculations show that the
hybrid functionals provide the most optimal one-particle orbitals,
which minimize the fixed-node error in the quantum Monte Carlo
method. The use of two independent codes and basis sets provides a
solid computational framework for accurate benchmarking of these
issues. In addition, study of high-pressure phase adds another
important test, since the dramatic changes in electronic structure
probe the pseudopotentials in the regime, for which their reliability
cannot be taken for granted.

\section{Ambient pressure}
\label{sec:ambient}

Before we proceed with investigation of the B1 $\to$ B8 phase
transition we first concentrate on energy differences between various
magnetic phases at ambient pressure. We apply the all-electron method
side by side with the pseudopotential approach. For this purpose we
have modified the WIEN2k package \cite{wien2k} so that one can use
either the original all-electron core solver or the norm-conserving
pseudopotentials (PP). The valence states are expanded in the LAPW
basis (extended by local orbitals) in both cases. This allows for a
systematic comparison of full cores with PPs, since the only point
where such calculations differ is the treatment of the core
electrons. Similar combination of LAPW and pseudopotentials was
already used to look into some aspects of PP applications to
transition metals \cite{cho1996}.

In order to minimize errors associated with partitioning core-valence
correlations in pseudopotential calculations, we have chosen
pseudopotentials with $3s$ and $3p$ states in the valence space, so
that the eliminated core corresponded to the Ne atom. We tested two
types of norm-conserving PPs constructed from Dirac-Fock atomic
solutions. The first one, denoted as YN, is a soft pseudopotential
generated using Troullier-Martins construction \cite{lee_private}. The
second PP, labeled as STU, is the so-called energy-adjusted
pseudopotential,\cite{dolg1987} which has the $-Z_{eff}/r$ attractive
ionic part retained.

In addition to performance of pseudopotentials we were interested in
the impact of exact exchange on electronic structure of manganese
oxide.  To this end we have employed the CRYSTAL2003 code
\cite{crystal2003} that uses Gaussian type atomic orbitals as a basis
for one-particle wave functions. In the following we refer to this
method as LCAO. The exact exchange was studied by means of two hybrid
exchange-correlation functionals. We utilized B3LYP as well as
somewhat simpler PBE0 functional. The later one is schematically
written as
\begin{equation}
E^{PBE0}_{xc}=a E_x^{HF} + (1-a) E^{PBE}_{x} + E^{PBE}_c\,,
\end{equation}
where $E^{PBE}_{x}$ and $E^{PBE}_c$ are exchange and correlation parts
of the PBE-GGA \cite{perdew1996}, $E_x^{HF}$ is the exact (Fock)
exchange and $a>0$. This type of hybrid mixing has the advantage that
it directly relates to the PBE-GGA approximation employed in our LAPW
calculations. We use a rather small value $a=1/10$ for the Fock mixing
weight to account for the (nearly) metallic behavior at pressures we
are going to investigate later. In the rest of the paper we abbreviate
this functional as PBE0${}_{10}$. With the standard choice of $a=1/4$
(Ref.~~\onlinecite{perdew1996b}) we ran into serious difficulties in
converging the Kohn-Sham equations when the lattice was compressed to
the extent needed to reach the B1 $\to$ B8 transition. Another
alternative and possibly more sophisticated way around this problem is
to employ a screened exchange instead of the full one
\cite{kasinathan2006,heyd2003}.

\begin{table}
\begin{tabular}{|l|>{$\,$}l<{$\,$}|>{$\,\,}c<{\,\,$
  \vrule height 10pt width 0pt}|>{$}c<{$}|>{$\,\,}c<{\,\,$}|}
\hline
\multirow{2}{*}{XC functional} & \multirow{2}{*}{core}
  & \Delta & \Delta^{all}\!-\!\Delta^{PP} & m \\
& & \hbox{(eV)}    & \hbox{(eV)}  & \hbox{($\mu_B$)}   \\
\hline\hline
\multirow{3}{*}{GGA, LAPW} &
all-electron & -2{.}29  &         & 4{.}23 \\
&YN          & -2{.}80  &  0{.}50 & 4{.}29 \\
&STU         & -2{.}94  &  0{.}65 & 4{.}32 \\
\hline
\multirow{3}{*}{\vbox{\hbox{GGA, LAPW,}\vskip .3em\hbox{non-relativistic}}} &
all-electron & -2{.}31  &         & 4{.}23 \\
&YN          & -2{.}78  &  0{.}47 & 4{.}29 \\
&STU         & -2{.}92  &  0{.}61 & 4{.}31 \\
\hline
\multirow{3}{*}{GGA, LCAO} &
all-electron & -2{.}39  &         & 4{.}59 \\
&YN          & -2{.}83  &  0{.}44 & 4{.}63 \\
&STU         & -2{.}96  &  0{.}57 & 4{.}64 \\
\hline
\multirow{3}{*}{\vbox{\hbox{PBE0${}_{10}$,}\vskip .3em\hbox{LCAO}}} &
all-electron & -3{.}50  &         & 4{.}70 \\
&YN          & -3{.}82  & 0{.}33  & 4{.}71 \\
&STU         & -3{.}91  & 0{.}41  & 4{.}71 \\
\hline
\multirow{3}{*}{B3LYP, LCAO} &
all-electron & -4{.}15  &         & 4{.}74 \\
&YN          & -4{.}24  &  0{.}09 & 4{.}74 \\
&STU         & -4{.}33  &  0{.}18 & 4{.}74 \\
\hline
\multirow{3}{*}{HF, LCAO} &
all-electron & -12{.}73 &         & 4{.}93 \\
&YN          & -12{.}55 & -0{.}18 & 4{.}91 \\
&STU         & -12{.}18 & -0{.}55 & 4{.}90 \\
\hline
\end{tabular}
\caption{\label{tab:MnO_errors} Energy difference
  $\Delta=E_{AFM}-E_{NM}$ per MnO at lattice
  constant $4{.}43\,$\AA\ together with local moments $m$ formed on Mn
  atoms in the AFM phase. The disagreement between 
  $\Delta^{all}$ and $\Delta^{PP}$ has opposite signs in PBE-GGA 
  and Hartee-Fock (HF) calculations and is considerably reduced for 
  hybrid functionals, especially for B3LYP.}
\end{table}

The outcome of the electronic structure methods introduced above is
summarized in Tab.~\ref{tab:MnO_errors} where we compare two
electronic phases of MnO in the rocksalt structure at equilibrium
lattice constant: the nonmagnetic (NM) state and the state with
antiferromagnetic (AFM) ordering of adjacent Mn (111)-planes, i.e.,
the so-called type-II antiferromagnet. The AFM phase represents the
ambient pressure ground state. In the case of the LAPW method, the
basis size and the density of $k$-point mesh was converged down to 1
or 2 mRy ($\sim 0{.}03$ eV) in obtained total energies per formula
unit. The Gaussian basis set for the LCAO methods was chosen large
enough to reproduce the (nonrelativistic) LAPW/GGA results (see
Tab.~\ref{tab:MnO_errors}).

Since we utilized only small-core pseudopotentials we expected only
minor discrepancies between the all-electron and PP calculations.
However, this assumption turned out not to be entirely correct. The
results listed in Tab.~\ref{tab:MnO_errors} show that both {\itshape
explicit presence of the core} and {\itshape the method which is
employed} do matter. The difference between the all-electron and
pseudopotential techniques varies from $+0{.}6$ eV/MnO within the
PBE-GGA functional to $-0{.}6$~eV/MnO in the Hartree-Fock method
combined with the STU PP. While contributions of this magnitude might
not be always relevant, for the purpose of comparing energies of
different magnetic phases or in high pressure transitions such a
difference turns out to be very significant.

In addition to the antiferromagnetic and nonmagnetic B1 phases we
included into our investigation also the ferromagnetically (FM)
ordered B1 structure. This time the energy difference
$\Delta=E_{AFM}-E_{FM}$ stays virtually the same irrespective of the
the core treatment used. This indicates that the sensitivity of
$\Delta$ to the description of the core electrons is related to
difference in magnetic moments of the two involved states --- the
local moments in FM and AFM phases are almost identical, approximately
$4{.}5\,\mu_B$ (see Tab.~\ref{tab:MnO_errors}), whereas they are by
definition zero in the NM state. More detailed evidence pointing in
this direction is presented within the high pressure study bellow.

Several comparisons of all-electron and small-core pseudopotential
data can be found in the literature in connection with the so-called
non-linear core correction (NLCC) \cite{louie1982}. LDA studies of
3$d$ atoms\cite{porezag1999} and transition metal bearing diatomic
molecules\cite{engel2001a} show that without the NLCC the errors in
energetics, associated with small-core pseudopotentials, are generally
of the order of tenths of eV, which is in agreement with our
observations made in solids. The non-linear core correction is
demonstrated to be capable of removing a large portion of these
errors\cite{porezag1999,engel2001a}.

We have decided not the include the NLCC in our study for the
following reason. The NLCC concept, correcting for linearization of
core-valence correlations inherent to the standard norm-conserving
pseudopotential technique, is not systematically transferable to
methods using exact exchange, i.e., to hybrid exchange-correlation
functionals, or to many-body methods such as quantum Monte Carlo. For
these approaches it is therefore important to assess how accurately
the actual pseudopotential Hamiltonian represents the physical
ion. Indeed, our study is a test whether this appears to be true for a
variety of magnetic states, range of pressures and different
methods. In the next section we provide additional benchmarks for MnO
molecule and Mn atom, which clearly delineate the consistency of the
used PPs and also suggest what is the reason for the discrepancies we
found. Although the discussion of pseudopotential constructions is out
of the scope of this paper we just note that the core-valence
partitioning is more straightforward within the Dirac-Fock
approximation than within the DFT due to the rather complicated
nonlinear dependence of the DFT exchange-correlation functional on the
density.  In fact, one set of the Dirac-Fock pseudopotentials we
employed has been used in high-accuracy correlated calculations by
quantum-chemical methods in a routine manner without additional
corrections and with satisfactory results for equilibrium properties
of molecular systems \cite{dolg1987b,wagner2003,wagner2006}.

\section{Testing the pseudopotentials in M\lowercase{n} atom and
  M\lowercase{n}O molecule}

\begin{table}
\begin{tabular}{|l|>{$\,$}l<{$\,$}|>{$\,\,}c<{\,\,$
  \vrule height 10pt width 0pt}|>{$}c<{$}|}
\hline
\multirow{2}{*}{method} & \multirow{2}{*}{core}
  & D_e         & D_e^{all}\!-\!D_e^{PP} \\
& & \hbox{(eV)} & \hbox{(eV)}            \\
\hline\hline
\multirow{3}{*}{ROHF} &
all-electron & -0{.}86  &          \\
&YN          & -0{.}85  & -0{.}01  \\
&STU         & -0{.}94  & \phantom{-}0{.}08  \\
\hline
\multirow{3}{*}{UHF} &
all-electron & \phantom{-}0{.}69  &          \\
&YN          & \phantom{-}0{.}66  & \phantom{-}0{.}03  \\
&STU         & \phantom{-}0{.}60  & \phantom{-}0{.}09  \\
\hline
\end{tabular}
\caption{\label{tab:molecule_dissoc}Dissociation energy $D_e$ of MnO
  molecule at a bond length $1{.}64\,$\AA\ calculated with restricted
  open shell (ROHF) and unrestricted (UHF) Hartree-Fock methods. Both
  the molecule and Mn atom are in a high-spin configuration with spin
  multiplicity~$6$.}
\end{table}

\begin{table}
\begin{tabular}{|l|>{$\,$}l<{$\,$}|>{$\,\,}c<{\,\,$
  \vrule height 10pt width 0pt}|>{$}c<{$}|}
\hline
\multirow{2}{*}{method} & \multirow{2}{*}{core}
  & \Delta      & \Delta^{all}\!-\!\Delta^{PP} \\
& & \hbox{(eV)} & \hbox{(eV)}            \\
\hline\hline
\multirow{3}{*}{\vbox{\hbox{$E_{ls}$ with ROHF}%
\vskip.3em\hbox{$E_{hs}$ with ROHF}}} &
all-electron & 6{.}59  &          \\
&YN          & 6{.}69  & -0{.}10  \\
&STU         & 6{.}76  & -0{.}17  \\
\hline
\multirow{3}{*}{\vbox{\hbox{$E_{ls}$ with ROHF}%
\vskip.3em\hbox{$E_{hs}$ with UHF}}} &
all-electron & 8{.}42  &          \\
&YN          & 8{.}50  & -0{.}08  \\
&STU         & 8{.}61  & -0{.}19  \\
\hline
\end{tabular}
\caption{\label{tab:molecule_hsls}Energy difference
  $\Delta=E_{ls}-E_{hs}$ between high-spin
  ($\sigma^1\delta^1\bar\delta^1\pi^1\bar\pi^1$, multiplicity 6) and
  low-spin ($\sigma^1\delta^2\bar\delta^2$, multiplicity 2)
  configurations of MnO molecule calculated 
  with Hartree-Fock methods. Bond length is $1{.}64\,$\AA\ as in
  Tab.~\ref{tab:molecule_dissoc}.}
\end{table}

The results presented in the preceding paragraph might appear somewhat
surprising, since very small errors generated by small-core PPs have
been observed in isolated atoms \cite{dolg1987,wagner2003}. In
particular, our Hartree-Fock tests of atomic $s\to d$ transfer
energies, i.e., differences between high-spin and low-spin states in
the isolated atom (not explicitly shown), turned out the same no
matter whether all-electron or pseudopotential Hamiltonian has been
employed. In addition, our previous calculations within high-accuracy
quantum Monte Carlo reproduced the experimental $s\to d$ transfer
energy within 0.1 eV \cite{wagner2003}. This suggests that the
norm-conserving Dirac-Fock construction provides accurate PP
Hamiltonian, which represents the physical ion within both the
mean-field and correlated many-body methods with accuracy better than
0.1 eV. A question then arises where is the origin of the
discrepancies reported in the previous section.  Is it something
related solely to the solid state environment or can it be detected as
soon as chemical bonds are formed? The LDA study of molecules
containing transition metal atoms, Ref.~~\onlinecite{engel2001a},
suggests that the second statement is indeed true. To verify this we
performed several Hartree-Fock calculations to demonstrate that such a
conclusion is not only LDA specific.

Tables~\ref{tab:molecule_dissoc} and~\ref{tab:molecule_hsls} show
results obtained using the GAMESS code \cite{gamess} with large
uncontracted basis sets for pseudoatoms and with natural atomic
orbitals\cite{roos1stbasis,roosTMbasis} for all-electron
atoms. Whenever only same spin states are involved, as is the case of
the dissociation energy $D_e$ (Tab.~\ref{tab:molecule_dissoc}) of the
ground state of MnO molecule, the all-electron and pseudopotential
results agree quite well.  Especially when YN pseudopotential is used,
the difference $D_e^{all} - D_e^{PP}$ falls quite clearly bellow
resolution of the basis sets employed. In addition, tests done with
highly correlated multi-determinant Slater-Jastrow trial wave
functions and with fixed-node diffusion Monte Carlo reproduced the
experimental binding energy within 0.1 eV
\cite{wagner2006}. Therefore, even cohesive properties of the PP
Hamiltonian look correct both within mean-field and many-body
correlated methods.

A small but visible inconsistency between all-electron and
pseudopotential approaches suddenly appears if one considers a
{\itshape high-spin to low-spin transition in the molecule}
(Tab.~\ref{tab:molecule_hsls}). Electronic configurations are
$\sigma^1\delta^1\bar\delta^1\pi^1\bar\pi^1$ for the high-spin and
$\sigma^1\delta^2\bar\delta^2$ for the low-spin state. Comparison of
quantities $\Delta^{all} - \Delta^{PP}$ between
Tabs.~\ref{tab:MnO_errors} and~\ref{tab:molecule_hsls} shows that the
discrepancies in the solid are almost three times larger than in the
molecule. This is understandable since the Mn bonding in solid is
enhanced by the octahedral arrangement of the oxygen atoms.  Trying to
estimate how much stronger this effect would be from the size of
cohesions one finds that the ratio of (experimental) binding energies
\cite{barin1993,smoes1984} of the two systems is
$9{.}5\,\hbox{eV}/3{.}8\,\hbox{eV} = 2{.}5\,$. The key question which
therefore remains to be answered is: where is this effect coming from?
We believe that the bonding environment is the key consideration for
resolving this issue. In an isolated atom the $d$~electrons have
freedom to expand away from the nucleus and indeed one finds that the
atomic $d$~orbitals for low-spin states are very diffuse. In the
presence of chemical bond, say, in a molecule, such an expansion is
not favorable, since bonding tends to localize the states even for
doubly occupied $d$~channels. In a solid this is even more pronounced
as the Mn atom is encapsulated in the octahedron of oxygens. This
enhances the amount of charge in the $d$~states inside the core. The
cause for the spin related discrepancies then becomes apparent --- it
is the bonding (or chemical) ``pressure''.

\section{Solid at high pressures (GGA)}

\begin{table}
\begin{tabular}{|l|>{$\,}c<{\,$\vrule height 10pt width 0pt}|
                 >{$\,}c<{\,$}|>{$\,}c<{\,$}|>{$\,}c<{\,$}|>{$\,}c<{\,$}|}
\hline
\multirow{2}{*}{method} & P_c & V_{B1}     & V_{B8}  & m_{B1}  & m_{B8}\\
              & \hbox{(GPa)} & \hbox{(\AA${}^3$)} & \hbox{(\AA$^3$)}
              & \hbox{($\mu_B$)} & \hbox{($\mu_B$)}  \\
\hline\hline
GGA, all-el. LAPW            &  43 & 18{.}15 & 15{.}76 & 3{.}8 & 2{.}3 \\
GGA, YN LAPW                 &  67 & 17{.}01 & 14{.}98 & 3{.}9 & 2{.}4 \\
PBE0${}_{10}$, all-el. LCAO  & 117 & 15{.}48 & 13{.}71 & 4{.}4 & 2{.}0 \\
PBE0${}_{10}$, YN LCAO       & 125 & 15{.}12 & 13{.}17 & 4{.}4 & 2{.}0 \\
\hline
\end{tabular}
\caption{\label{tab:pressures}Critical pressures for the B1 AFM $\to$
  B8 FM phase transition as calculated with various methods. Also
  indicated are volumes (per formula unit) and the local magnetic
  moments on Mn atoms for both phases at the transition.
}
\end{table}

In order to
 examine the high pressure behavior of MnO, namely the B1 $\to$ B8
transition, we have carried out a series of LAPW/GGA calculations%
\endnote{The muffin-tin radii $R_{MT}$ were set to $1.81\, a_B$ for
  Mn atom and $1.6\, a_B$ for oxygen ($a_B$ stands
  for Bohr radius) and held constant for all volumes. This choice
  represents almost touching spheres for the smallest studied volume.
}
for the type-II antiferromagnetic phase in the B1 structure and for
the B8 structure with ferromagnetic as well as antiferromagnetic
ordering of hexagonal Mn planes.%
\endnote{In the B8 phases the Mn atoms occupy the Ni sites and oxygen
atoms sit at the As sites.}
The critical pressure~$P_c$ and volume change at the B1 $\to$ B8
transition are listed on the first two lines of
Tab.~\ref{tab:pressures}. The arrangement of local moments is found to
change from AFM to FM across the transition. The choice of the
magnetic ordering in the high pressure phase is not well resolved,
however, since the energy difference between B8 AFM and B8 FM phases
at the relevant volumes is comparable to our 2~mRy precision
estimate. The Fermi level lies in the gap between bands of the
Kohn-Sham eigenvalues from ambient pressure all the way up to the
transition in the case of the B1 AFM phase, whereas it is located
inside a band for both studied B8 phases above and also well bellow
$P_c\,$. These qualitative features do not change if full cores or
pseudopotentials are used. The variation in quantitative aspects is,
on the other hand, quite sizeable. For instance, the difference in the
transition pressure~$P_c$ is as large as 25 GPa.

The GGA predictions are rather far from experimental $P_c$ around
100~GPa \cite{kondo2000,yoo2005,patterson2004} and from a non-magnetic
high pressure phase \cite{yoo2005} observed at room temperature (our
calculations would support a paramagnetic phase at this temperature,
at best).  This is not surprising as the GGA is not expected to
approximate the electronic structure of MnO very accurately. At the
same time, the pressure $P_c=67$~GPa resulting from our PP
calculations is in a good agreement with results of Fang {\itshape et
al.}~\cite{fang1999} who obtained $P_c\approx 60$~GPa with a
plane-wave pseudopotential approach. The remaining few GPa might
originate in the full structure optimization performed in
Ref.~~\onlinecite{fang1999}. For the sake of simplicity, we have
neglected the rhombohedral distortion in the B1 AFM phase and have not
optimized the $c/a$ ratio in the B8 phases. Instead we used a constant
value $c/a=2.0$ consistent with outcome of Ref.~~\onlinecite{fang1999}
and slightly smaller than $c/a\approx 2.1$ suggested by x-ray
spectroscopy \cite{yoo2005}.

\begin{figure}
\resizebox{\linewidth}{!}{\includegraphics{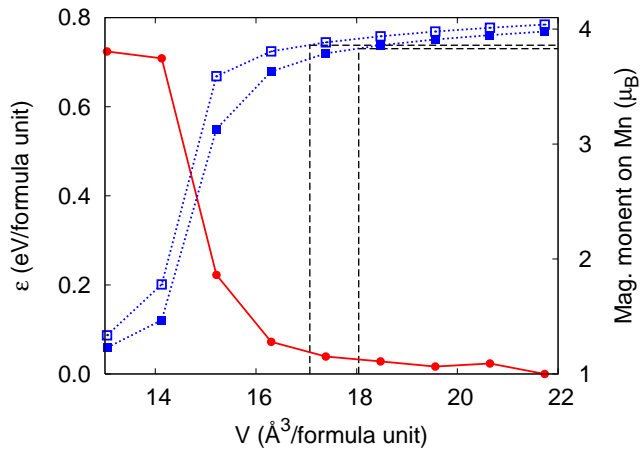}}
\caption{\label{fig:ecp_error_B1}(color online) Difference between
  all-electron and pseudopotential total energies,
  Eq.~\eqref{eq:AE_PP_difference}, as calculated for the B1 AFM
  structure within LAPW/GGA method (bullets). Also shown is magnetic
  moment on the Mn atom  with all-electron core (full squares) and YN
  pseudopotential (empty squares). The slight enhancement of local
  moments when pseudopotentials are introduced, visible also in
  Tabs.~\ref{tab:MnO_errors} and~\ref{tab:pressures}, is consistent
  with other reports \cite{cocula2003,cho1996}. Volumes at the
  transition to the B8 FM phase are indicated as well.}
\end{figure}

\begin{figure}
\resizebox{\linewidth}{!}{\includegraphics{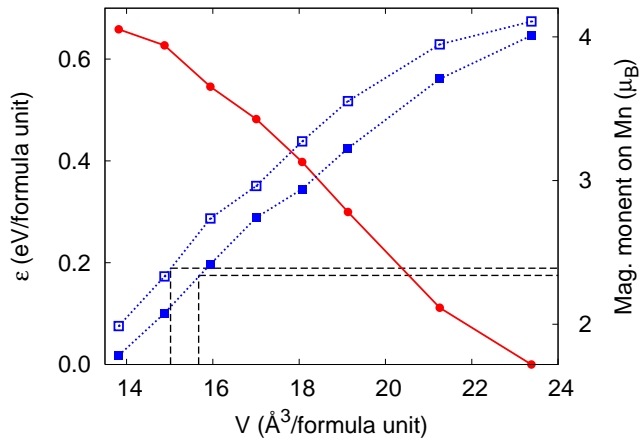}}
\caption{\label{fig:ecp_error_B8}(color online) Same quantities as in
  Fig.~\ref{fig:ecp_error_B1} but for B8 FM structure.}
\end{figure}

Figures~\ref{fig:ecp_error_B1} and~\ref{fig:ecp_error_B8} relate the
discrepancy between the two distinct core electron treatments to
magnetic moments formed on manganese atoms. The quantity we use to
measure the difference between all-electron and pseudopotential total
energies is defined as
\begin{equation}\label{eq:AE_PP_difference}
\varepsilon(V)=E^{all}(V)-E^{PP}(V)-\bigl[E^{all}(V_0)-E^{PP}(V_0)\bigr]\,,
\end{equation}
where $V_0$ stands for a reference volume. We use
$V_0=21{.}7\,$\AA${}^3$ per formula unit in the B1 phase
(Fig.~\ref{fig:ecp_error_B1}) and $V_0=23{.}4\,$\AA${}^3$ per formula
unit in the B8 phase (Fig.~\ref{fig:ecp_error_B8}). In the rocksalt
structure, the difference~$\varepsilon$ remains almost constant as
long as the spin polarization does not change, and jumps up at the
high-spin to low-spin transition \cite{cohen1997}. The main reason for
the 25 GPa inconsistency in the critical pressure~$P_c$ lies in the B8
phase, however, since rather high pressure is necessary for the
high-spin to low-spin transition in the rocksalt structure.  In the B8
phase the magnetic moment decreases gradually with volume compression
and this reduction is accompanied with a gradual change of the
difference~$\varepsilon$ that reaches an appreciable value at the
point of the phase transition (see Fig.~\ref{fig:ecp_error_B8}). The
fact that the variation of the total energy difference~$\varepsilon$
apparently follows the evolution of spin polarization suggests that
some type of spin-dependent pseudopotentials, possibly those
introduced in Ref.~~\onlinecite{watson1998}, might bring the
all-electron and pseudopotential results closer together.

\section{Solid at high pressures (hybrid functional)}

\begin{figure}
\resizebox{\linewidth}{!}{\includegraphics{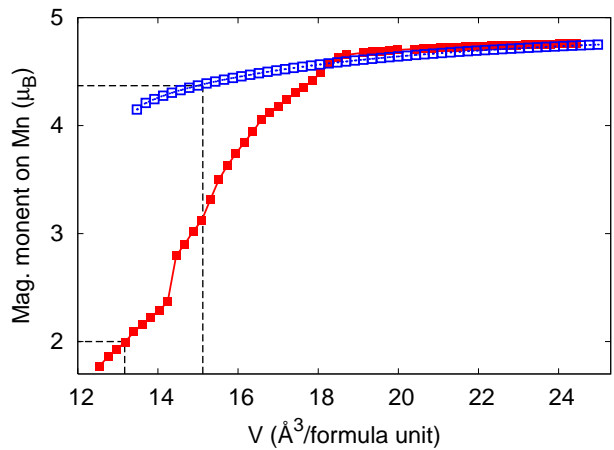}}
\caption{\label{fig:hybrid_moments}(color online) Magnetic moment on
  Mn atom in B1 AFM (empty squares) and B8 FM (full squares) phases as
  calculated using the PBE0${}_{10}$ functional. Values are taken from
  YN PP calculation, corresponding all-electron data are virtually
  indistinguishable. Transition volumes are shown as well.} 
\end{figure}

To evaluate how the exact exchange influences the critical pressure of
the B1 $\to$ B8 transition we repeated the calculations with the
PBE0${}_{10}$ functional as shown on the last two lines of
Table~\ref{tab:pressures}. We can see that already a small admixture
of exact exchange moves the critical pressure $P_c$ way above the GGA
prediction, just a little higher than experimentally observed.
Corresponding volumes of the phases above and bellow the transition
are also very close to the experimental results \cite{yoo2005}.  We
would like to emphasize, however, that this study is targeted on
methodology rather than on precise comparison with experiments since
the latter goal would certainly require detailed geometry
optimizations. The overall picture of the transition within
PBE0${}_{10}$ stays the same as with the GGA, the rocksalt phase is an
insulator, the B8 phase is a metal. The magnetic moments collapse from
$4{.}4\,\mu_B$ to $2.0\,\mu_B$ as the MnO is compressed across the
transition. (Note that there are slight inconsistencies in comparisons
of LAPW and LCAO local magnetic moments due to their different
definitions.)  The evolution of the magnetic moments with decreasing
volume is shown in Fig.~\ref{fig:hybrid_moments}. The tendency of the
Fock exchange towards electron localization is clearly visible, the
decrease in moments is delayed to a lower volume in both B1 and B8
phases as compared to the GGA results displayed in
Figs.~\ref{fig:ecp_error_B1} and~\ref{fig:ecp_error_B8}. In the
rocksalt structure we actually see no high-spin to low-spin transition
in the interval of examined volumes.

The discrepancy between all-electron and pseudopotential approaches is
greatly reduced when a hybrid functional is used. Following the
discussion of the GGA results one might expect that this reduction is
accompanied by limited spin changes. However, the
Fig.~\ref{fig:hybrid_moments} demonstrates that this not the case and
therefore we conclude that the core treated by PBE0${}_{10}$ is more
inert and is less sensitive to changes of the local moment formed by
valence electrons.

\section{Conclusions}

Our findings complement several other reports found in the literature
indicating that application of pseudopotential based techniques to
transition metals and their compounds is a delicate task
\cite{kresse1999}. In particular, we can mention discrepancies in
magnetic properties of the vanadium surface
\cite{batyrev2001,kresse2002,cocula2004} or inconsistencies in
predictions of the ground state of Ce${}_2$O${}_3$
\cite{kresse2005,fabris2005b} found between pseudopotential and
all-electron methods. In these studies, however, the large-core (Ar)
PPs were utilized. The difficulties with pseudopotentials are then
attributed to the semi-core $3s$ and $3p$ states that have too large
overlap with the valence $4s$ and $3d$ electrons to be accurately
represented by a ``rigid core''. In the same time, the semi-core
states are usually too localized to be efficiently included in the
valence space.%
\endnote{The problem of a large (plane-wave) cut-off required by
  localized semi-core states can be successfully overcome within
  alternative frozen-core techniques such as ultrasoft
  pseudopotentials \cite{vanderbilt1990} or the projector
  augmented-wave (PAW) method.\cite{blochl1994,kresse1999}{}}
This type of argument, however, cannot explain the variations in the
B1 $\to$ B8 transition pressure presented in this paper, since we
treat the $3s$ and $3p$ states as valence degrees of freedom. Our
observations suggest that one should carefully examine transferability
of pseudopotentials between different spin states even in the case of
small-core PPs, especially if the phenomena under investigation
involve variations in local magnetic moments.

Overall implications of our results are rather straightforward to
explain. The $3d$ states have no counterparts of the same symmetry in
the core and therefore an appreciable amount of the $d$ charge density
is localized in the core region. Consequently, changes in the local
moments can influence the core states and vice versa.  As we have
verified, this is not detectable in isolated atoms, where the
all-electron and pseudopotential results agree very well, nor in
equilibrium ground state cohesion energies, which do not involve
dramatic changes in the local moment on the Mn atom. This is true not
only for the mean-field methods employed in this study, but also for
highly accurate quantum Monte Carlo tests done previously
\cite{wagner2003,wagner2006}. These results suggest that the PP
Hamiltonian is consistent and accurate. The discrepancy appears and
becomes visible when the {\itshape high-spin/low-spin differences} become
involved in bonds. The localization of the $d$ states due to the
chemical environment pushes larger fraction of the charge into the
core and inaccuracies of various methods become more pronounced. The
direction of the biases, which we found, is also not very difficult to
understand. The LDA and GGA functionals are well-known to incline
towards more smooth and homogeneous densities with exaggerated
softening of electron-electron repulsion. On the opposite side is the
Hartree-Fock method, which tends to localize the states since that
enhances the exchange contributions (the only mechanism, through which
HF can decrease the energy). The key point is that these biases affect
{\itshape all degrees of freedom} of the given system, including, in
all-electron calculations, also the core states. Therefore, the cores
are ``softened'' in the LDA/GGA approaches and ``tightened'' in the HF
approximation.  In the PP case the core is rigid and therefore the
bias from core states is absent that leads to seemingly better (closer
to experiment) results for the transition pressures.  Finally, these
results could be of interest not only for high-pressure solids but
also for other transition metal systems with significant magnetic
moment changes such as biomolecular reaction centers or catalysis.

\acknowledgments

We acknowledge support by NSF DMR-0121361 and EAR-0530110 grants and
we would like to thank for the kind consent to modify the WIEN2k code
by P.~Blaha.

%\bibliography{tmo,electronic_structure_methods,pseudopotentials}

\begin{thebibliography}{40}
\expandafter\ifx\csname natexlab\endcsname\relax\def\natexlab#1{#1}\fi
\expandafter\ifx\csname bibnamefont\endcsname\relax
  \def\bibnamefont#1{#1}\fi
\expandafter\ifx\csname bibfnamefont\endcsname\relax
  \def\bibfnamefont#1{#1}\fi
\expandafter\ifx\csname citenamefont\endcsname\relax
  \def\citenamefont#1{#1}\fi
\expandafter\ifx\csname url\endcsname\relax
  \def\url#1{\texttt{#1}}\fi
\expandafter\ifx\csname urlprefix\endcsname\relax\def\urlprefix{URL }\fi
\providecommand{\bibinfo}[2]{#2}
\providecommand{\eprint}[2][]{\url{#2}}

\bibitem[{\citenamefont{Mazin et~al.}(1998)\citenamefont{Mazin, Fei, Dawns, and
  Cohen}}]{mazin1998}
\bibinfo{author}{\bibfnamefont{I.~I.} \bibnamefont{Mazin}},
  \bibinfo{author}{\bibfnamefont{Y.}~\bibnamefont{Fei}},
  \bibinfo{author}{\bibfnamefont{R.}~\bibnamefont{Dawns}}, \bibnamefont{and}
  \bibinfo{author}{\bibfnamefont{R.}~\bibnamefont{Cohen}},
  \bibinfo{journal}{American Mineralogist} \textbf{\bibinfo{volume}{83}},
  \bibinfo{pages}{451} (\bibinfo{year}{1998}).

\bibitem[{\citenamefont{Fang et~al.}(1998)\citenamefont{Fang, Terakura, Sawada,
  Miyazaki, and Solovyev}}]{fang1998}
\bibinfo{author}{\bibfnamefont{Z.}~\bibnamefont{Fang}},
  \bibinfo{author}{\bibfnamefont{K.}~\bibnamefont{Terakura}},
  \bibinfo{author}{\bibfnamefont{H.}~\bibnamefont{Sawada}},
  \bibinfo{author}{\bibfnamefont{T.}~\bibnamefont{Miyazaki}}, \bibnamefont{and}
  \bibinfo{author}{\bibfnamefont{I.}~\bibnamefont{Solovyev}},
  \bibinfo{journal}{Phys. Rev. Lett.} \textbf{\bibinfo{volume}{81}},
  \bibinfo{pages}{1027} (\bibinfo{year}{1998}).

\bibitem[{\citenamefont{Pask et~al.}(2001)\citenamefont{Pask, Singh, Mazin,
  Hellberg, and Kortus}}]{pask2001}
\bibinfo{author}{\bibfnamefont{J.~E.} \bibnamefont{Pask}},
  \bibinfo{author}{\bibfnamefont{D.~J.} \bibnamefont{Singh}},
  \bibinfo{author}{\bibfnamefont{I.~I.} \bibnamefont{Mazin}},
  \bibinfo{author}{\bibfnamefont{C.~S.} \bibnamefont{Hellberg}},
  \bibnamefont{and} \bibinfo{author}{\bibfnamefont{J.}~\bibnamefont{Kortus}},
  \bibinfo{journal}{Phys. Rev. B} \textbf{\bibinfo{volume}{64}},
  \bibinfo{pages}{024403} (\bibinfo{year}{2001}).

\bibitem[{\citenamefont{Terakura et~al.}(1984)\citenamefont{Terakura, Oguchi,
  Williams, and K\"ubler}}]{terakura1984}
\bibinfo{author}{\bibfnamefont{K.}~\bibnamefont{Terakura}},
  \bibinfo{author}{\bibfnamefont{T.}~\bibnamefont{Oguchi}},
  \bibinfo{author}{\bibfnamefont{A.~R.} \bibnamefont{Williams}},
  \bibnamefont{and} \bibinfo{author}{\bibfnamefont{J.}~\bibnamefont{K\"ubler}},
  \bibinfo{journal}{Phys. Rev. B} \textbf{\bibinfo{volume}{30}},
  \bibinfo{pages}{4734} (\bibinfo{year}{1984}).

\bibitem[{\citenamefont{Kasinathan et~al.}()\citenamefont{Kasinathan, {Kune\v
  s}, Koepernik, Diaconu, Martin, Prodan, Scuseria, Spaldin, Petit, Schulthess
  et~al.}}]{kasinathan2006}
\bibinfo{author}{\bibfnamefont{D.}~\bibnamefont{Kasinathan}},
  \bibinfo{author}{\bibfnamefont{J.}~\bibnamefont{{Kune\v s}}},
  \bibinfo{author}{\bibfnamefont{K.}~\bibnamefont{Koepernik}},
  \bibinfo{author}{\bibfnamefont{C.~V.} \bibnamefont{Diaconu}},
  \bibinfo{author}{\bibfnamefont{R.~L.} \bibnamefont{Martin}},
  \bibinfo{author}{\bibfnamefont{I.}~\bibnamefont{Prodan}},
  \bibinfo{author}{\bibfnamefont{G.~E.} \bibnamefont{Scuseria}},
  \bibinfo{author}{\bibfnamefont{N.}~\bibnamefont{Spaldin}},
  \bibinfo{author}{\bibfnamefont{L.}~\bibnamefont{Petit}},
  \bibinfo{author}{\bibfnamefont{T.~C.} \bibnamefont{Schulthess}},
  \bibnamefont{and} \bibinfo{author}{\bibfnamefont{W.~E.}
  \bibnamefont{Pickett}}, \bibinfo{note}{cond-mat/0605430}.

\bibitem[{\citenamefont{Cohen et~al.}(1997)\citenamefont{Cohen, Mazin, and
  Isaak}}]{cohen1997}
\bibinfo{author}{\bibfnamefont{R.~E.} \bibnamefont{Cohen}},
  \bibinfo{author}{\bibfnamefont{I.~I.} \bibnamefont{Mazin}}, \bibnamefont{and}
  \bibinfo{author}{\bibfnamefont{D.~G.} \bibnamefont{Isaak}},
  \bibinfo{journal}{Science} \textbf{\bibinfo{volume}{275}},
  \bibinfo{pages}{654} (\bibinfo{year}{1997}).

\bibitem[{\citenamefont{Kondo et~al.}(2000)\citenamefont{Kondo, Yagi, Syono,
  Noguchi, Atou, Kikegawa, and Shimomura}}]{kondo2000}
\bibinfo{author}{\bibfnamefont{T.}~\bibnamefont{Kondo}},
  \bibinfo{author}{\bibfnamefont{T.}~\bibnamefont{Yagi}},
  \bibinfo{author}{\bibfnamefont{Y.}~\bibnamefont{Syono}},
  \bibinfo{author}{\bibfnamefont{Y.}~\bibnamefont{Noguchi}},
  \bibinfo{author}{\bibfnamefont{T.}~\bibnamefont{Atou}},
  \bibinfo{author}{\bibfnamefont{T.}~\bibnamefont{Kikegawa}}, \bibnamefont{and}
  \bibinfo{author}{\bibfnamefont{O.}~\bibnamefont{Shimomura}},
  \bibinfo{journal}{J. Appl. Phys.} \textbf{\bibinfo{volume}{87}},
  \bibinfo{pages}{4153} (\bibinfo{year}{2000}).

\bibitem[{\citenamefont{Yoo et~al.}(2005)\citenamefont{Yoo, Maddox, Klepeis,
  Iota, Evans, McMahan, Hu, Chow, Somayazulu, Hausermann et~al.}}]{yoo2005}
\bibinfo{author}{\bibfnamefont{C.~S.} \bibnamefont{Yoo}},
  \bibinfo{author}{\bibfnamefont{B.}~\bibnamefont{Maddox}},
  \bibinfo{author}{\bibfnamefont{J.-H.~P.} \bibnamefont{Klepeis}},
  \bibinfo{author}{\bibfnamefont{V.}~\bibnamefont{Iota}},
  \bibinfo{author}{\bibfnamefont{W.}~\bibnamefont{Evans}},
  \bibinfo{author}{\bibfnamefont{A.}~\bibnamefont{McMahan}},
  \bibinfo{author}{\bibfnamefont{M.~Y.} \bibnamefont{Hu}},
  \bibinfo{author}{\bibfnamefont{P.}~\bibnamefont{Chow}},
  \bibinfo{author}{\bibfnamefont{M.}~\bibnamefont{Somayazulu}},
  \bibinfo{author}{\bibfnamefont{D.}~\bibnamefont{Hausermann}},
  \bibinfo{author}{\bibfnamefont{R.~T.} \bibnamefont{Scalettar}},
  \bibnamefont{and} \bibinfo{author}{\bibfnamefont{W.~E.}
  \bibnamefont{Pickett}}, \bibinfo{journal}{Phys. Rev. Lett.}
  \textbf{\bibinfo{volume}{94}}, \bibinfo{eid}{115502} (\bibinfo{year}{2005}).

\bibitem[{\citenamefont{Patterson et~al.}(2004)\citenamefont{Patterson, Aracne,
  Jackson, Malba, Weir, Baker, and Vohra}}]{patterson2004}
\bibinfo{author}{\bibfnamefont{J.~R.} \bibnamefont{Patterson}},
  \bibinfo{author}{\bibfnamefont{C.~M.} \bibnamefont{Aracne}},
  \bibinfo{author}{\bibfnamefont{D.~D.} \bibnamefont{Jackson}},
  \bibinfo{author}{\bibfnamefont{V.}~\bibnamefont{Malba}},
  \bibinfo{author}{\bibfnamefont{S.~T.} \bibnamefont{Weir}},
  \bibinfo{author}{\bibfnamefont{P.~A.} \bibnamefont{Baker}}, \bibnamefont{and}
  \bibinfo{author}{\bibfnamefont{Y.~K.} \bibnamefont{Vohra}},
  \bibinfo{journal}{Phys. Rev. B} \textbf{\bibinfo{volume}{69}},
  \bibinfo{eid}{220101(R)} (\bibinfo{year}{2004}).

\bibitem[{\citenamefont{Wagner and Mitas}(2003)}]{wagner2003}
\bibinfo{author}{\bibfnamefont{L.}~\bibnamefont{Wagner}} \bibnamefont{and}
  \bibinfo{author}{\bibfnamefont{L.}~\bibnamefont{Mitas}},
  \bibinfo{journal}{Chem. Phys. Lett.} \textbf{\bibinfo{volume}{370}},
  \bibinfo{pages}{412} (\bibinfo{year}{2003}).

\bibitem[{\citenamefont{Wagner and Mitas}()}]{wagner2006}
\bibinfo{author}{\bibfnamefont{L.~K.} \bibnamefont{Wagner}} \bibnamefont{and}
  \bibinfo{author}{\bibfnamefont{L.}~\bibnamefont{Mitas}},
  \bibinfo{note}{cond-mat/0610088}.

\bibitem[{\citenamefont{Louie et~al.}(1982)\citenamefont{Louie, Froyen, and
  Cohen}}]{louie1982}
\bibinfo{author}{\bibfnamefont{S.~G.} \bibnamefont{Louie}},
  \bibinfo{author}{\bibfnamefont{S.}~\bibnamefont{Froyen}}, \bibnamefont{and}
  \bibinfo{author}{\bibfnamefont{M.~L.} \bibnamefont{Cohen}},
  \bibinfo{journal}{Phys. Rev. B} \textbf{\bibinfo{volume}{26}},
  \bibinfo{pages}{1738} (\bibinfo{year}{1982}).

\bibitem[{\citenamefont{Cho and Scheffler}(1996)}]{cho1996}
\bibinfo{author}{\bibfnamefont{J.-H.} \bibnamefont{Cho}} \bibnamefont{and}
  \bibinfo{author}{\bibfnamefont{M.}~\bibnamefont{Scheffler}},
  \bibinfo{journal}{Phys. Rev. B} \textbf{\bibinfo{volume}{53}},
  \bibinfo{pages}{10685} (\bibinfo{year}{1996}).

\bibitem[{\citenamefont{Kresse and Joubert}(1999)}]{kresse1999}
\bibinfo{author}{\bibfnamefont{G.}~\bibnamefont{Kresse}} \bibnamefont{and}
  \bibinfo{author}{\bibfnamefont{D.}~\bibnamefont{Joubert}},
  \bibinfo{journal}{Phys. Rev. B} \textbf{\bibinfo{volume}{59}},
  \bibinfo{pages}{1758} (\bibinfo{year}{1999}).

\bibitem[{\citenamefont{Kiejna et~al.}(2006)\citenamefont{Kiejna, Kresse,
  Rogal, {De Sarkar}, Reuter, and Scheffler}}]{kiejna2006}
\bibinfo{author}{\bibfnamefont{A.}~\bibnamefont{Kiejna}},
  \bibinfo{author}{\bibfnamefont{G.}~\bibnamefont{Kresse}},
  \bibinfo{author}{\bibfnamefont{J.}~\bibnamefont{Rogal}},
  \bibinfo{author}{\bibfnamefont{A.}~\bibnamefont{{De Sarkar}}},
  \bibinfo{author}{\bibfnamefont{K.}~\bibnamefont{Reuter}}, \bibnamefont{and}
  \bibinfo{author}{\bibfnamefont{M.}~\bibnamefont{Scheffler}},
  \bibinfo{journal}{Phys. Rev. B} \textbf{\bibinfo{volume}{73}},
  \bibinfo{eid}{035404} (\bibinfo{year}{2006}).

\bibitem[{\citenamefont{Blaha et~al.}(2001)\citenamefont{Blaha, Schwarz,
  Madsen, Kvasnicka, and Luitz}}]{wien2k}
\bibinfo{author}{\bibfnamefont{P.}~\bibnamefont{Blaha}},
  \bibinfo{author}{\bibfnamefont{K.}~\bibnamefont{Schwarz}},
  \bibinfo{author}{\bibfnamefont{G.~K.~H.} \bibnamefont{Madsen}},
  \bibinfo{author}{\bibfnamefont{D.}~\bibnamefont{Kvasnicka}},
  \bibnamefont{and} \bibinfo{author}{\bibfnamefont{J.}~\bibnamefont{Luitz}},
  \emph{\bibinfo{title}{WIEN2k, An Augmented Plane Wave + Local Orbitals
  Program for Calculating Crystal Properties}} (\bibinfo{publisher}{Techn.
  {Universit\"at} Wien, Austria}, \bibinfo{year}{2001}).

\bibitem[{\citenamefont{Lee}()}]{lee_private}
\bibinfo{author}{\bibfnamefont{Y.}~\bibnamefont{Lee}}, \bibinfo{note}{private
  communication}.

\bibitem[{\citenamefont{Dolg et~al.}(1987{\natexlab{a}})\citenamefont{Dolg,
  Wedig, Stoll, and Preuss}}]{dolg1987}
\bibinfo{author}{\bibfnamefont{M.}~\bibnamefont{Dolg}},
  \bibinfo{author}{\bibfnamefont{U.}~\bibnamefont{Wedig}},
  \bibinfo{author}{\bibfnamefont{H.}~\bibnamefont{Stoll}}, \bibnamefont{and}
  \bibinfo{author}{\bibfnamefont{H.}~\bibnamefont{Preuss}},
  \bibinfo{journal}{J. Chem. Phys.} \textbf{\bibinfo{volume}{86}},
  \bibinfo{pages}{866} (\bibinfo{year}{1987}{\natexlab{a}}).

\bibitem[{\citenamefont{Saunders et~al.}(2003)\citenamefont{Saunders, Dovesi,
  Roetti, Orlando, Zicovich-Wilson, Harrison, Doll, Civalleri, Bush, D'Arco
  et~al.}}]{crystal2003}
\bibinfo{author}{\bibfnamefont{V.}~\bibnamefont{Saunders}},
  \bibinfo{author}{\bibfnamefont{R.}~\bibnamefont{Dovesi}},
  \bibinfo{author}{\bibfnamefont{C.}~\bibnamefont{Roetti}},
  \bibinfo{author}{\bibfnamefont{R.}~\bibnamefont{Orlando}},
  \bibinfo{author}{\bibfnamefont{C.~M.} \bibnamefont{Zicovich-Wilson}},
  \bibinfo{author}{\bibfnamefont{N.~M.} \bibnamefont{Harrison}},
  \bibinfo{author}{\bibfnamefont{K.}~\bibnamefont{Doll}},
  \bibinfo{author}{\bibfnamefont{B.}~\bibnamefont{Civalleri}},
  \bibinfo{author}{\bibfnamefont{I.}~\bibnamefont{Bush}},
  \bibinfo{author}{\bibfnamefont{P.}~\bibnamefont{D'Arco}}, \bibnamefont{and}
  \bibinfo{author}{\bibfnamefont{M.}~\bibnamefont{Llunell}},
  \emph{\bibinfo{title}{CRYSTAL2003 User's Manual}}
  (\bibinfo{publisher}{University of Torino, Torino}, \bibinfo{year}{2003}).

\bibitem[{\citenamefont{Perdew et~al.}(1996{\natexlab{a}})\citenamefont{Perdew,
  Burke, and Ernzerhof}}]{perdew1996}
\bibinfo{author}{\bibfnamefont{J.~P.} \bibnamefont{Perdew}},
  \bibinfo{author}{\bibfnamefont{K.}~\bibnamefont{Burke}}, \bibnamefont{and}
  \bibinfo{author}{\bibfnamefont{M.}~\bibnamefont{Ernzerhof}},
  \bibinfo{journal}{Phys. Rev. Lett.} \textbf{\bibinfo{volume}{77}},
  \bibinfo{pages}{3865} (\bibinfo{year}{1996}{\natexlab{a}}).

\bibitem[{\citenamefont{Perdew et~al.}(1996{\natexlab{b}})\citenamefont{Perdew,
  Ernzerhof, and Burke}}]{perdew1996b}
\bibinfo{author}{\bibfnamefont{J.~P.} \bibnamefont{Perdew}},
  \bibinfo{author}{\bibfnamefont{M.}~\bibnamefont{Ernzerhof}},
  \bibnamefont{and} \bibinfo{author}{\bibfnamefont{K.}~\bibnamefont{Burke}},
  \bibinfo{journal}{J. Chem. Phys.} \textbf{\bibinfo{volume}{105}},
  \bibinfo{pages}{9982} (\bibinfo{year}{1996}{\natexlab{b}}).

\bibitem[{\citenamefont{Heyd et~al.}(2003)\citenamefont{Heyd, Scuseria, and
  Ernzerhof}}]{heyd2003}
\bibinfo{author}{\bibfnamefont{J.}~\bibnamefont{Heyd}},
  \bibinfo{author}{\bibfnamefont{G.~E.} \bibnamefont{Scuseria}},
  \bibnamefont{and}
  \bibinfo{author}{\bibfnamefont{M.}~\bibnamefont{Ernzerhof}},
  \bibinfo{journal}{J. Chem. Phys.} \textbf{\bibinfo{volume}{118}},
  \bibinfo{pages}{8207} (\bibinfo{year}{2003}).

\bibitem[{\citenamefont{Porezag et~al.}(1999)\citenamefont{Porezag, Pederson,
  and Liu}}]{porezag1999}
\bibinfo{author}{\bibfnamefont{D.}~\bibnamefont{Porezag}},
  \bibinfo{author}{\bibfnamefont{M.~R.} \bibnamefont{Pederson}},
  \bibnamefont{and} \bibinfo{author}{\bibfnamefont{A.~Y.} \bibnamefont{Liu}},
  \bibinfo{journal}{Phys. Rev. B} \textbf{\bibinfo{volume}{60}},
  \bibinfo{pages}{14132} (\bibinfo{year}{1999}).

\bibitem[{\citenamefont{Engel et~al.}(2001)\citenamefont{Engel, H\"ock, and
  Varga}}]{engel2001a}
\bibinfo{author}{\bibfnamefont{E.}~\bibnamefont{Engel}},
  \bibinfo{author}{\bibfnamefont{A.}~\bibnamefont{H\"ock}}, \bibnamefont{and}
  \bibinfo{author}{\bibfnamefont{S.}~\bibnamefont{Varga}},
  \bibinfo{journal}{Phys. Rev. B} \textbf{\bibinfo{volume}{63}},
  \bibinfo{pages}{125121} (\bibinfo{year}{2001}).

\bibitem[{\citenamefont{Dolg et~al.}(1987{\natexlab{b}})\citenamefont{Dolg,
  Wedig, Stoll, and Preuss}}]{dolg1987b}
\bibinfo{author}{\bibfnamefont{M.}~\bibnamefont{Dolg}},
  \bibinfo{author}{\bibfnamefont{U.}~\bibnamefont{Wedig}},
  \bibinfo{author}{\bibfnamefont{H.}~\bibnamefont{Stoll}}, \bibnamefont{and}
  \bibinfo{author}{\bibfnamefont{H.}~\bibnamefont{Preuss}},
  \bibinfo{journal}{J. Chem. Phys.} \textbf{\bibinfo{volume}{86}},
  \bibinfo{pages}{2123} (\bibinfo{year}{1987}{\natexlab{b}}).

\bibitem[{\citenamefont{Schmidt et~al.}(1993)\citenamefont{Schmidt, Baldridge,
  Boatz, Elbert, Gordon, Jensen, Koseki, Matsunaga, Nguyen, Su
  et~al.}}]{gamess}
\bibinfo{author}{\bibfnamefont{M.~W.} \bibnamefont{Schmidt}},
  \bibinfo{author}{\bibfnamefont{K.~K.} \bibnamefont{Baldridge}},
  \bibinfo{author}{\bibfnamefont{J.~A.} \bibnamefont{Boatz}},
  \bibinfo{author}{\bibfnamefont{S.~T.} \bibnamefont{Elbert}},
  \bibinfo{author}{\bibfnamefont{M.~S.} \bibnamefont{Gordon}},
  \bibinfo{author}{\bibfnamefont{J.~H.} \bibnamefont{Jensen}},
  \bibinfo{author}{\bibfnamefont{S.}~\bibnamefont{Koseki}},
  \bibinfo{author}{\bibfnamefont{N.}~\bibnamefont{Matsunaga}},
  \bibinfo{author}{\bibfnamefont{K.~A.} \bibnamefont{Nguyen}},
  \bibinfo{author}{\bibfnamefont{S.~J.} \bibnamefont{Su}},
  \bibinfo{author}{\bibfnamefont{T.~L.} \bibnamefont{Windus}},
  \bibinfo{author}{\bibfnamefont{M.}~\bibnamefont{Dupuis}}, \bibnamefont{and}
  \bibinfo{author}{\bibfnamefont{J.~A.} \bibnamefont{Montgomery}},
  \bibinfo{journal}{J. Comput. Chem.} \textbf{\bibinfo{volume}{14}},
  \bibinfo{pages}{1347} (\bibinfo{year}{1993}).

\bibitem[{\citenamefont{Widmark et~al.}(1990)\citenamefont{Widmark, Malmqvist,
  and Roos}}]{roos1stbasis}
\bibinfo{author}{\bibfnamefont{P.-O.} \bibnamefont{Widmark}},
  \bibinfo{author}{\bibfnamefont{P.}~\bibnamefont{Malmqvist}},
  \bibnamefont{and} \bibinfo{author}{\bibfnamefont{B.~O.} \bibnamefont{Roos}},
  \bibinfo{journal}{Theor. Chim. Acta} \textbf{\bibinfo{volume}{77}},
  \bibinfo{pages}{291} (\bibinfo{year}{1990}).

\bibitem[{\citenamefont{{Pou-Am\' erigo} et~al.}(1995)\citenamefont{{Pou-Am\'
  erigo}, {Merch\' an}, Nebot-Gil, Widmark, and Roos}}]{roosTMbasis}
\bibinfo{author}{\bibfnamefont{R.}~\bibnamefont{{Pou-Am\' erigo}}},
  \bibinfo{author}{\bibfnamefont{M.}~\bibnamefont{{Merch\' an}}},
  \bibinfo{author}{\bibfnamefont{I.}~\bibnamefont{Nebot-Gil}},
  \bibinfo{author}{\bibfnamefont{P.-O.} \bibnamefont{Widmark}},
  \bibnamefont{and} \bibinfo{author}{\bibfnamefont{B.~O.} \bibnamefont{Roos}},
  \bibinfo{journal}{Theor. Chim. Acta} \textbf{\bibinfo{volume}{92}},
  \bibinfo{pages}{149} (\bibinfo{year}{1995}).

\bibitem[{\citenamefont{Barin}(1993)}]{barin1993}
\bibinfo{author}{\bibfnamefont{I.}~\bibnamefont{Barin}},
  \emph{\bibinfo{title}{Thermochemical data of pure substances}}
  (\bibinfo{publisher}{VCH, Weinheim, Federal Republic of Germany},
  \bibinfo{year}{1993}), \bibinfo{edition}{2nd} ed.

\bibitem[{\citenamefont{Smoes and Drowart}(1984)}]{smoes1984}
\bibinfo{author}{\bibfnamefont{S.}~\bibnamefont{Smoes}} \bibnamefont{and}
  \bibinfo{author}{\bibfnamefont{J.}~\bibnamefont{Drowart}},
  \bibinfo{journal}{High Temp. Sci.} \textbf{\bibinfo{volume}{17}},
  \bibinfo{pages}{31} (\bibinfo{year}{1984}).

\bibitem[{\citenamefont{Fang et~al.}(1999)\citenamefont{Fang, Solovyev, Sawada,
  and Terakura}}]{fang1999}
\bibinfo{author}{\bibfnamefont{Z.}~\bibnamefont{Fang}},
  \bibinfo{author}{\bibfnamefont{I.~V.} \bibnamefont{Solovyev}},
  \bibinfo{author}{\bibfnamefont{H.}~\bibnamefont{Sawada}}, \bibnamefont{and}
  \bibinfo{author}{\bibfnamefont{K.}~\bibnamefont{Terakura}},
  \bibinfo{journal}{Phys. Rev. B} \textbf{\bibinfo{volume}{59}},
  \bibinfo{pages}{762} (\bibinfo{year}{1999}).

\bibitem[{\citenamefont{Cocula et~al.}(2003)\citenamefont{Cocula, Starrost,
  Watson, and Carter}}]{cocula2003}
\bibinfo{author}{\bibfnamefont{V.}~\bibnamefont{Cocula}},
  \bibinfo{author}{\bibfnamefont{F.}~\bibnamefont{Starrost}},
  \bibinfo{author}{\bibfnamefont{S.~C.} \bibnamefont{Watson}},
  \bibnamefont{and} \bibinfo{author}{\bibfnamefont{E.~A.}
  \bibnamefont{Carter}}, \bibinfo{journal}{J. Chem. Phys.}
  \textbf{\bibinfo{volume}{119}}, \bibinfo{pages}{7659} (\bibinfo{year}{2003}).

\bibitem[{\citenamefont{Watson and Carter}(1998)}]{watson1998}
\bibinfo{author}{\bibfnamefont{S.~C.} \bibnamefont{Watson}} \bibnamefont{and}
  \bibinfo{author}{\bibfnamefont{E.~A.} \bibnamefont{Carter}},
  \bibinfo{journal}{Phys. Rev. B} \textbf{\bibinfo{volume}{58}},
  \bibinfo{pages}{R13309} (\bibinfo{year}{1998}).

\bibitem[{\citenamefont{Batyrev et~al.}(2001)\citenamefont{Batyrev, Cho, and
  Kleinman}}]{batyrev2001}
\bibinfo{author}{\bibfnamefont{I.~G.} \bibnamefont{Batyrev}},
  \bibinfo{author}{\bibfnamefont{J.-H.} \bibnamefont{Cho}}, \bibnamefont{and}
  \bibinfo{author}{\bibfnamefont{L.}~\bibnamefont{Kleinman}},
  \bibinfo{journal}{Phys. Rev. B} \textbf{\bibinfo{volume}{63}},
  \bibinfo{pages}{172420} (\bibinfo{year}{2001}).

\bibitem[{\citenamefont{Kresse et~al.}(2002)\citenamefont{Kresse, Bergermayer,
  and Podloucky}}]{kresse2002}
\bibinfo{author}{\bibfnamefont{G.}~\bibnamefont{Kresse}},
  \bibinfo{author}{\bibfnamefont{W.}~\bibnamefont{Bergermayer}},
  \bibnamefont{and}
  \bibinfo{author}{\bibfnamefont{R.}~\bibnamefont{Podloucky}},
  \bibinfo{journal}{Phys. Rev. B} \textbf{\bibinfo{volume}{66}},
  \bibinfo{pages}{146401} (\bibinfo{year}{2002}).

\bibitem[{\citenamefont{Cocula and Carter}(2004)}]{cocula2004}
\bibinfo{author}{\bibfnamefont{V.}~\bibnamefont{Cocula}} \bibnamefont{and}
  \bibinfo{author}{\bibfnamefont{E.~A.} \bibnamefont{Carter}},
  \bibinfo{journal}{Phys. Rev. B} \textbf{\bibinfo{volume}{69}},
  \bibinfo{eid}{052404} (\bibinfo{year}{2004}).

\bibitem[{\citenamefont{Kresse et~al.}(2005)\citenamefont{Kresse, Blaha, {Da
  Silva}, and Ganduglia-Pirovano}}]{kresse2005}
\bibinfo{author}{\bibfnamefont{G.}~\bibnamefont{Kresse}},
  \bibinfo{author}{\bibfnamefont{P.}~\bibnamefont{Blaha}},
  \bibinfo{author}{\bibfnamefont{J.~L.~F.} \bibnamefont{{Da Silva}}},
  \bibnamefont{and} \bibinfo{author}{\bibfnamefont{M.~V.}
  \bibnamefont{Ganduglia-Pirovano}}, \bibinfo{journal}{Phys. Rev. B}
  \textbf{\bibinfo{volume}{72}}, \bibinfo{eid}{237101} (\bibinfo{year}{2005}).

\bibitem[{\citenamefont{Fabris et~al.}(2005)\citenamefont{Fabris, de~Gironcoli,
  Baroni, Vicario, and Balducci}}]{fabris2005b}
\bibinfo{author}{\bibfnamefont{S.}~\bibnamefont{Fabris}},
  \bibinfo{author}{\bibfnamefont{S.}~\bibnamefont{de~Gironcoli}},
  \bibinfo{author}{\bibfnamefont{S.}~\bibnamefont{Baroni}},
  \bibinfo{author}{\bibfnamefont{G.}~\bibnamefont{Vicario}}, \bibnamefont{and}
  \bibinfo{author}{\bibfnamefont{G.}~\bibnamefont{Balducci}},
  \bibinfo{journal}{Phys. Rev. B} \textbf{\bibinfo{volume}{72}},
  \bibinfo{eid}{237102} (\bibinfo{year}{2005}).

\bibitem[{\citenamefont{Vanderbilt}(1990)}]{vanderbilt1990}
\bibinfo{author}{\bibfnamefont{D.}~\bibnamefont{Vanderbilt}},
  \bibinfo{journal}{Phys. Rev. B} \textbf{\bibinfo{volume}{41}},
  \bibinfo{pages}{7892} (\bibinfo{year}{1990}).

\bibitem[{\citenamefont{Bl\"ochl}(1994)}]{blochl1994}
\bibinfo{author}{\bibfnamefont{P.~E.} \bibnamefont{Bl\"ochl}},
  \bibinfo{journal}{Phys. Rev. B} \textbf{\bibinfo{volume}{50}},
  \bibinfo{pages}{17953} (\bibinfo{year}{1994}).

\end{thebibliography}

\end{document}